# Signatures of the differential Klein-Nishina electronic cross section in Compton's quantum theory of scattering of radiation

Vinay Venugopal* and Piyush S Bhagdikar

Division of Physics, School of Advanced Sciences, VIT University, Chennai Campus, Vandalur Kelambakkam Road, Chennai-600127, India

*Corresponding author, e-mail: vinayvenugopal@vit.ac.in

**A quantum theory of scattering of radiation by a stationary free electron based on photon conception[1] and relativistic kinematics, applying the principles of conservation of energy and conservation of momentum was proposed by Compton[2] (independently by Debye[3]) to explain the scattering of X-rays and γ-rays by light elements[2]. The relativistic differential cross-section $d\sigma/d\Omega_{\phi} \equiv KN(\phi)$ for the Compton scattering of a photon by a stationary free electron (where $\phi$ is the scattering angle of the photon) was formulated by Klein and Nishina[4] (independently by Tamm[5]) using Dirac's relativistic theory of electrons[6], and has been verified experimentally, when the binding energy of the electron is negligible compared to the incident photon energy ($h\nu$). Here we show that the energy of scattered photons, and kinetic energy of recoiled electrons obtained from Compton's quantum theory of scattering of radiation, show a degree of matching (that increases with the increase of $h\nu$ as quantified by chi-square test) with the**

**differential Klein-Nishina electronic cross section per electron per unit solid angle $KN(\phi)$ for the scattering of an unpolarized photon by a stationary free electron when appropriate normalizations are invoked. There is a high degree of matching in a regime where the total electronic Klein-Nishina cross section for the Compton scattering on a free stationary electron scales as $(h\nu)^{-1}$ and the contribution of the electro-magnetic interaction to $KN(\phi)$ diminishes. Our results have significant implications to the foundations of quantum mechanics and to the understanding of the mechanisms of photon-electron interactions in the Compton scattering.**

Our recent work[7] on the de Broglie wavelength and frequency of recoiled electrons in the Compton effect motivated us to investigate the possibilities of signature of $KN(\phi)$[4,5] in the energy of the scattered photons ($h\nu'(\phi)$), and momentum ($p_e'(\phi)$) and kinetic energy ($K_e'(\phi)$) of recoiled electrons respectively for $h\nu$ in the range of 0.00001-1000 MeV, obtained from the quantum theory proposed by Compton[2] to explain scattering of X-rays and γ–rays by light elements[2]. Appropriate normalizations are invoked to compare $KN(\phi)$ with $h\nu'(\phi)$, $p_e'(\phi)$, and $K_e'(\phi)$ respectively for a given $h\nu$, and it is shown that that the comparison with $h\nu'(\phi)$ and $K_e'(\phi)$ is physically meaningful.

The differential Klein-Nishina electronic cross section[4] per electron per unit solid angle for an unpolarized photon scattered at an angle $\phi$ by a free stationary electron is

$$\frac{d_e\sigma_C^{KN}}{d\Omega_\phi} = KN(\phi) = \frac{r_e^2}{2}\left(\frac{h\nu'}{h\nu}\right)^2\left(\frac{h\nu'}{h\nu} + \frac{h\nu}{h\nu'} - \sin^2\phi\right) \quad (1)$$

where $r_e$ is the classical radius of the electron, $\nu$ is the frequency of the incident photon, $\nu'$ and $\phi$ are the frequency and scattering angle of the scattered photon respectively (see Supplementary Fig. S1). For $h\nu \geq 1\text{MeV}$, the contribution of electromagnetic interaction to $KN(\phi)$ is maximum at $\phi = 0$ and minimum at $\phi = \pi$. The value of $KN(0)$ is the same as Thompson differential cross section per electron per unit solid angle for a photon scattered at $\phi = 0$. The value of $\left(KN(0) - KN(\phi)_{\min}\right)$ increases with the increase in $h\nu$, becoming asymptotic to the x-axis at high $h\nu$ (≥ 100 MeV). Hence $\left(KN(0) - KN(\phi)_{\min}\right)$ values for $h\nu$ in range 0.00001-1000 MeV are divided by the value of $\left(KN(0) - KN(\phi)_{\min}\right)$ for 1000 MeV and these normalized values are shown in Fig. 1b (see Supplementary Table S1). Subtracting $KN(\phi)_{\min}$ from the left-hand side of the equation (1) and dividing it by $\left(KN(0) - KN(\phi)_{\min}\right)$ for the respective $h\nu$ normalizes the corresponding $KN(0)$ to 1. For $h\nu$ in the range 0.001-1000 MeV, $\left(KN(\phi) - KN(\phi)_{\min}\right)$ is divided by the value of $\left(KN(0) - KN(\phi)_{\min}\right)_{1000\text{MeV}}$ and these normalized curves are shown in Fig. 1a.

In the quantum theory of scattering of X-rays by light elements, Compton considered the scattering of photon by a free stationary electron[2]. The energy of the scattered photon is related to the energy of the incident photon (see Supplementary Fig. S1)

$$h\nu'(\phi) = \frac{h\nu}{1 + \left(\dfrac{h\nu}{m_0 c^2}\right)(1 - \cos\phi)} \tag{2}$$

where $m_0$ is the rest mass of the electron. The momentum of the scattered photon is $p_{ph}{}'(\phi) = h\nu'(\phi)/c$. The values of $\left(h\nu'(0) - h\nu'(\pi)\right)$ for $h\nu$ in the range 1-1000 MeV are divided by the respective value of $h\nu$ and these normalized values are shown in Fig. 1d (see Supplementary Table S1). For incident photon energies less than 1 MeV, the $\pi$ in $h\nu'(\pi)$ is replaced by $\phi$ at which the corresponding $KN(\phi)$ is minimum. Note that dividing $\left(h\nu'(\phi)-h\nu'(\pi)\right)$ by $\left(h\nu'(0)-h\nu'(\pi)\right)$ for the respective $h\nu$ normalizes the corresponding $h\nu'(\phi)$ to 1. $h\nu'(\phi)$ is divided by the corresponding $h\nu$ (Compton scatter fraction) and these normalized curves are shown in Fig. 1c. For a given $h\nu$, $KN(\phi)$ and $h\nu'(\phi)$ respectively are normalized to 1 and their difference is plotted as the function of scattering angle of the scattered photon. These are shown for 0.001, 0.1, 1, and 1000 MeV in Figs. 2a, 2d, 2g and 2j respectively (see Supplementary Fig. S2 for incident photon energies of 0.01, 10, and 100 MeV respectively).With the increase in $h\nu$, the values of this difference approach the straight line $y = 0$. Identical results will be obtained for the momentum of the scattered photon.

The momentum of recoiled electrons $p_e'(\phi)$ in the Compton effect is obtained from the conservation of momentum (see Supplementary Fig. S1)

$$p_e'(\phi) = \left(\left(\frac{h\nu}{c}\right)^2 + \left(\frac{h\nu'}{c}\right)^2 - \left(\frac{2h\nu h\nu' \cos\phi}{c^2}\right)\right)^{1/2} \tag{3}$$

The incident photon momentum ($h\nu/c$) for $h\nu$ in the range of 1-1000 MeV is divided by the respective value of $p_e'(\pi)$ and these normalized values are shown in Fig. 1f (see Supplementary Table S1). For incident photon energies less than 1 MeV, the $\pi$ in $p_e'(\pi)$ are replaced by $\phi$ at which the corresponding $KN(\phi)$ is minimum. Note that dividing the left-hand side of equation (3) by $p_e'(\pi)$ for the respective incident photon energy normalizes the corresponding $p_e'(\phi)$ to 1. In Fig. 1e, these normalized values are multiplied by the corresponding $\left(p_{ph}/p_e'(\pi)\right)$ for $h\nu$ in the range 0.001-1000 MeV. For a given $h\nu$, $KN(\phi)$ and $p_e'(\phi)$ are normalized to 1 and their sum is plotted as function of photon scattering angle. These are shown for 0.001, 0.1, 1, and 1000 MeV in Figs. 2b, 2e, 2h and 2k respectively (see Supplementary Fig. S2 for incident photon energies of 0.01, 10, and 100 MeV respectively). With the increase in $h\nu$ the sum values approach the straight line $y = 1$.

The kinetic energy $K_e'(\phi)$ of recoiled electrons (see Supplementary Fig. S1) is

$$K_e'(\phi) = h\nu - h\nu'(\phi) \tag{4}$$

For $h\nu$ in the range 1-1000 MeV, $K_e'(\phi)$ are divided by the respective value of $h\nu$ and these normalized values (Compton energy transfer fraction) are shown in Fig. 1h (see Supplementary Table S1). For incident photon energies less than 1 MeV, the $\pi$ in $K_e'(\pi)$ are replaced by $\phi$ at which the corresponding $KN(\phi)$ is minimum. Note that dividing the left-hand side of equation (4) by $K_e'(\pi)$ for the respective $h\nu$ normalizes the corresponding $K_e'(\phi)$ to 1. $K_e'(\phi)$ is divided by the corresponding $h\nu$ in the range 0.001-1000 MeV, and these normalized curves are shown in Fig. 1g . For a given $h\nu$, $KN(\phi)$ and $K_e'(\phi)$ are normalized to 1 and their sum is plotted as function of photon scattering angle. These are shown for 0.001, 0.1, 1, and 1000 MeV in Figs. 2c, 2f, 2i and 2l respectively (see Supplementary Fig. S2 for incident photon energies of 0.01, 10, and 100 MeV respectively). With the increase in $h\nu$ the sum values approach the straight line $y = 1$.

The total energy of the recoiled electrons $\left(E_e'(\phi) = \gamma m_0 c^2\right)$ is obtained from conservation of energy (see Supplementary Fig. S1)

$$E_e'(\phi) = h\nu - h\nu' + m_0 c^2 \tag{5}$$

$E_e'(\pi) - m_0 c^2$ for $h\nu$ in the range 0.001-1000 MeV is divided by the respective values of $h\nu$ and these 'normalized' values will be identical to the fifth column in Supplemnetary Table S1 $\left(K_e'(\pi)/h\nu\right)$. The variation of $E_e'(\pi) - m_0 c^2$ with $h\nu$ will be identical to Fig. 1g. Subtracting $m_0 c^2$ from left-hand side of equation (6) and dividing by the respective $h\nu$, a plot identical to $K_e'(\phi)$ as in Fig. 1e will be obtained.

The degree of matching between 'observed' and 'expected' values in Fig. 2 is numerically analyzed by applying Pearson's chi-square test

$$\chi^2 = \sum_{i=1}^{n} \left( \frac{(O_i - E_i)^2}{E_i} \right) \tag{6}$$

where the 'observed' values are $O_i = \left(h\nu'(\phi_i)\right)_{Norm}$, $O_i = \left[\left(KN(\phi_i)\right)_{Norm} + \left(p_e'(\phi_i)\right)_{Norm}\right]$, and $O_i = \left[\left(KN(\phi_i)\right)_{Norm} + \left(K_e'(\phi_i)\right)_{Norm}\right]$ for the 'expected' values $E_i = \left(KN(\phi_i)\right)_{Norm}$, $E_i = 1$, and $E_i = 1$ respectively. We consider 2000 data points and hence a degree of freedom equal to 1999 in the 0 to π interval on the x-axis, where the critical value is 1740.7049, for a probability level of 0.9999. The values of chi-square are shown in

Supplementary Table S2. It is clear that these chi-square values are less than the critical value and they decrease with the increase of incident photon energy. The above analysis was repeated by considering the case when $\sin^2\phi = 0$ in $KN(\phi)$, since $\sin^2\phi$ corresponds to the Coulomb scattering of electron by the electromagnetic field of the incident photon[8]. The corresponding degree of matching between 'observed' and 'expected' values are shown in Fig. 3 and the values of chi-square are shown in Supplementary Table S3.

The above results (see Supplementary Table S2) clearly indicate that for a given incident energy of the photon in the range 0.0001-1000 MeV, there is a degree of matching between $\left[\left(KN(\phi)\right)_{Norm}\right]$ and $\left[\left(h\nu'(\phi)\right)_{Norm}\right]$, $\left[\left(KN(\phi)\right)_{Norm} + \left(p_e'(\phi)\right)_{Norm}\right]$ and 1, and $\left[\left(KN(\phi)\right)_{Norm} + \left(K_e'(\phi)\right)_{Norm}\right]$ and 1 respectively. The degree of matching increases with the increase in the energy of the incident photon in all the three cases. In addition, there is a degree of matching between $\left(KN(0) - KN(\phi)_{\min}\right)/\left(KN(0) - KN(\phi)_{\min}\right)_{1000\text{MeV}}$ and $\left[\left(h\nu'(0) - h\nu'(\pi)\right)/h\nu\right]$, $\left(p_{ph}/p_e'(\pi)\right)$, and $\left(K_e'(\pi)/h\nu\right)$ respectively for $h\nu \geq 1\text{MeV}$ as indicated in Supplementary Table S1. The degree of matching between the 'observed' (Compton) and 'expected' (Klein-Nishina) values in Supplementary Table S1 for $h\nu \geq 1\text{MeV}$ is numerically analyzed using chi-square test (equation (6)). For $O_i = \left[\left(h\nu'(0) - h\nu'(\pi)\right)/h\nu\right]$ and $E_i = \left(KN(0) - KN(\phi)_{\min}\right)/\left(KN(0) - KN(\phi)_{\min}\right)_{1000\text{MeV}}$, $\chi^2 = 0.01025$ when degree of freedom = 7, critical value = 0.1528, for a probability level of 0.9999. Similarly, for $O_i = \left[p_{ph}/p_e'(\pi)\right]$ and $\left(KN(0) - KN(\pi)\right)/\left(KN(0) - KN(\pi)\right)_{1000\text{MeV}}$, $\chi^2 = 0.02286$. A closer look at Supplementary Table S1 for $h\nu \le 1\text{MeV}$ and Fig. 3 indicates that only a comparison of $KN(\phi)$ with $h\nu'(\phi)$ and $K_e'(\phi)$ respectively is physically meaningful. The arguments that we propose here for comparing $KN(\phi)$ and $K_e'(\phi)$ will be the same as that of comparing $KN(\phi)$ and $h\nu'(\phi)$. The increasing degree of matching between $\left(KN(0)-KN(\phi)_{\min}\right)/\left(KN(0)-KN(\phi)_{\min}\right)_{1000\text{MeV}}$ and $\left(K_e'(\pi)/h\nu\right)$ for $h\nu \ge 1\text{MeV}$ (see Supplementary Table S1) is attributed to the increasing role of the contribution of energy-momentum conservation to $KN(\phi)$ at $\phi = \pi$ compared to the electromagnetic interaction. For $h\nu \le 1\text{MeV}$ (see Supplementary Table S1), the $\pi$ in $K_e'(\pi)$ are replaced by $\phi$ at which the corresponding $KN(\phi)$ is minimum. With the decrease in energy of the incident photon (see Supplementary Table S1), for a given $h\nu$, $\left(K_e'(\phi)/h\nu\right)$ values become much less than the corresponding $\left(KN(0)-KN(\phi)_{\min}\right)/\left(KN(0)-KN(\phi)_{\min}\right)_{1000\text{MeV}}$ which is attributed to the decreasing role of the energy-momentum conservation compared to electromagnetic interaction to the scattering at $\phi$ for which $KN(\phi)$ is minimum. Comparison of $KN(\phi)$ with $h\nu'(\phi)$ and $K_e'(\phi)$ respectively when $\sin^2\phi = 0$ in $KN(\phi)$ are shown in Figure 3 (see also Supplementary Table S3). It is clear that there is a high degree of matching between $\left[\left(KN(\phi)\right)_{Norm}\right]$ and $\left[\left(h\nu'(\phi)\right)_{Norm}\right]$, and $\left[\left(KN(\phi)\right)_{Norm} + \left(K_e'(\phi)\right)_{Norm}\right]$ and 1 at the lowest and highest incident photon energies when Coulomb scattering of electron by the electromagnetic field of the incident photon is ignored in $KN(\phi)$.

Therefore under the *conditions* when $\left(h\nu/m_0c^2\right)\geq 1.956$ , the signatures of differential Klein-Nishina cross section[4] for unpolarized photons that was obtained from Dirac's relativistic quantum mechanics[6] can be 'observed' in the energy of the scattered photons, and in the kinetic energy of recoiled electrons obtained from Compton's quantum theory of scattering of photon by electron based on relativistic kinematics[2]. Hence the third level explanation[9] of Compton effect by quantum electrodynamics has a degree of matching with the first level[9] of Compton[2]. These *conditions* also coincide with the total electronic Klein-Nishina cross section $\sigma_C^{KN}$ for the Compton scattering on a free stationary electron having a $(h\nu)^{-1}$ dependence, where the role of the electromagnetic interaction to the scattering process diminishes. Hence our results have significant implications to the foundations of quantum mechanics and to the understanding of the mechanisms of photon-electron interactions in the Compton scattering.

**Acknowledgements**

We thank Dr. S. Hariharan, Dr. S. G. Chefranov, and Dr. A. Datta for fruitful discussions. V.V thanks VIT University, Chennai Campus for providing a stimulating environment for innovative teaching and research infrastructure, and Dr. A. A. Samuel for his encouragement.

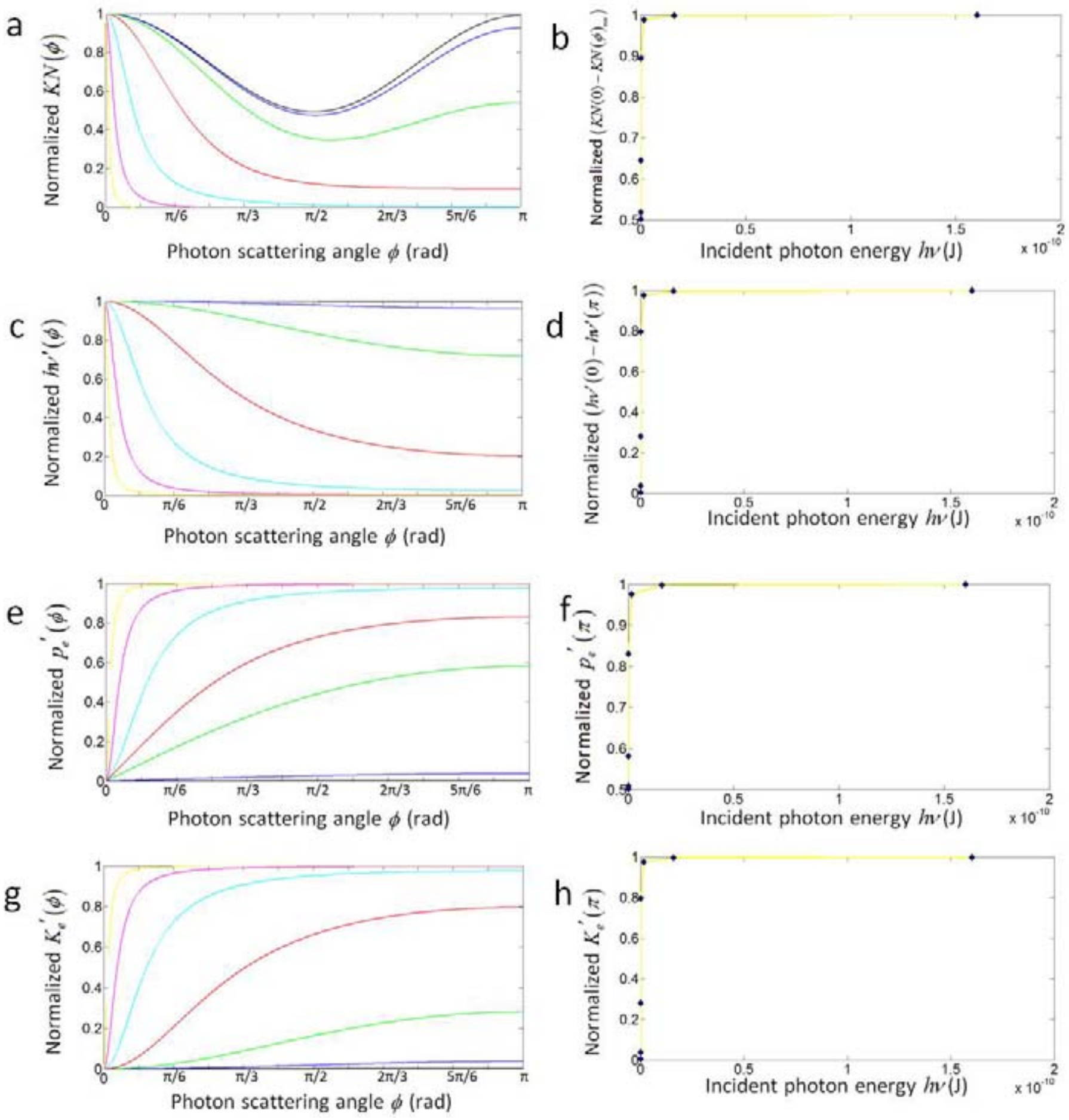

**Figure 1| Normalization**. The variation of normalized **a.** $KN(\phi)$, **c.** $h\nu'(\phi)$, **e.** $p_e'(\phi)$ and **g.** $K_e'(\phi)$ respectively as a function of the photon scattering angle $\phi$ for incident photon energies in the range 0.001-1000 MeV. Color code for incident photon energies: black-0.001 MeV, blue-0.01 MeV, green-0.1 MeV, red-1 MeV, cyan-10 MeV, magenta-100 MeV, yellow-1000 MeV. Variation of normalized **b.** $\left(KN(0)-KN(\phi)_{\min}\right)$, **d.** $\left(h\nu'(0)-h\nu'(\pi)\right)$, **f.** $p_e'(\pi)$ and **h.** $K_e'(\pi)$ respectively as a function of the photon scattering angle $\phi$ for incident photon energies in the range 1-1000 MeV. For incident photon energies less than 1 MeV, the $\pi$ in $h\nu'(\pi)$, $p_e'(\pi)$, and $K_e'(\pi)$ are replaced by $\phi$ at which the corresponding $KN(\phi)$ is minimum in **b**, **d**, **f**, and **h** respectively.

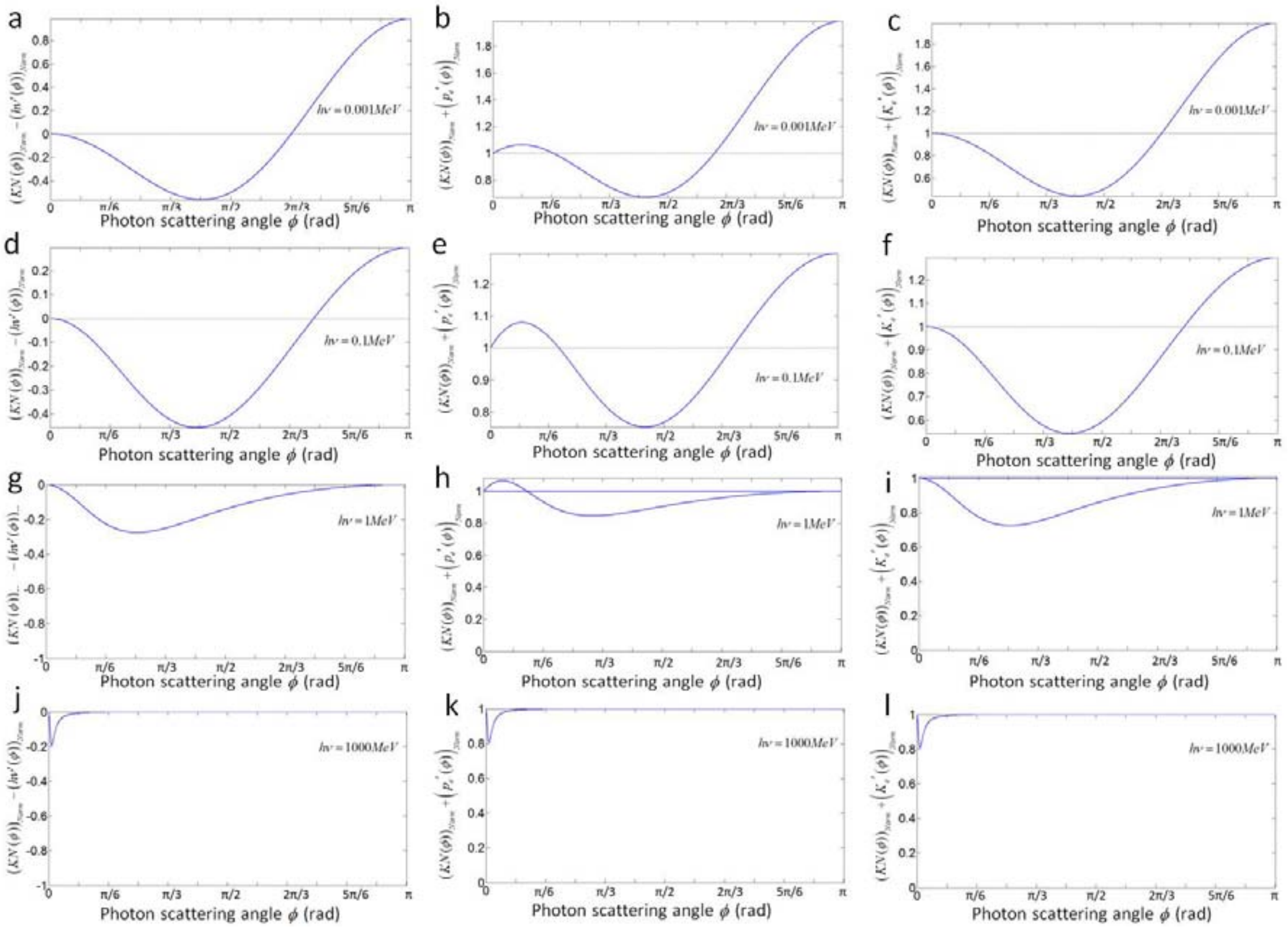

**Figure 2| Degree of matching.** The variation of $\left[\left(KN(\phi)\right)_{Norm} - \left(h\nu'(\phi)\right)_{Norm}\right]$ for incident photon energies **a.** 0.001 MeV, **d.** 0.1 MeV, **g.** 1 MeV, and **j.** 1000 MeV respectively, $\left[\left(KN(\phi)\right)_{Norm} + \left(p_e'(\phi)\right)_{Norm}\right]$ for **b.** 0.001 MeV, **e.** 0.1 MeV, **h.** 1 MeV, and **k.** 1000 MeV respectively, and $\left[\left(KN(\phi)\right)_{Norm} + \left(K_e'(\phi)\right)_{Norm}\right]$ for **c.** 0.001 MeV, **f.** 0.1 MeV, **i.** 1 MeV, and **l.** 1000 MeV respectively as a function of photon scattering angle $\phi$.

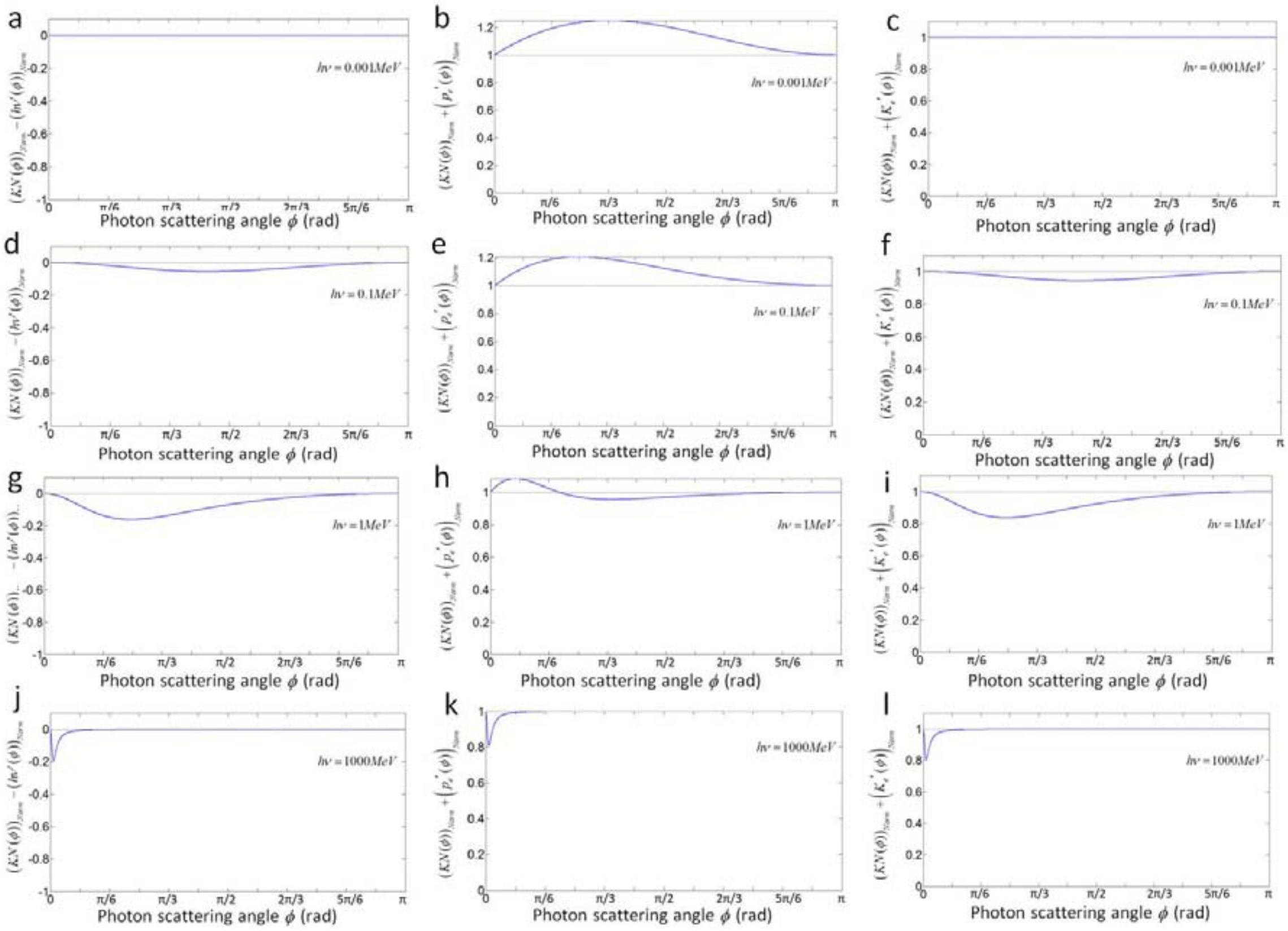

**Figure 3| Degree of matching when** $\sin^2\phi = 0$ **in** $KN(\phi)$. The variation of $\left[(KN(\phi))_{Norm} - (h\nu'(\phi))_{Norm}\right]$ for incident photon energies **a.** 0.001 MeV, **d.** 0.1 MeV, **g.** 1 MeV, and **j.** 1000 MeV respectively, $\left[(KN(\phi))_{Norm} + \left(p_e'(\phi)\right)_{Norm}\right]$ for **b.** 0.001 MeV, **e.** 0.1 MeV, **h.** 1 MeV, and **k.** 1000 MeV respectively, and $\left[(KN(\phi))_{Norm} + \left(K_e'(\phi)\right)_{Norm}\right]$ for **c.** 0.001 MeV, **f.** 0.1 MeV, **i.** 1 MeV, and **l.** 1000 MeV respectively as a function of photon scattering angle $\phi$.

## SUPPLEMENTARY INFORMATION

The values of the universal constants used in the numerical calculations in this Letter were obtained from NIST database http://physics.nist.gov/cuu/Constants/.

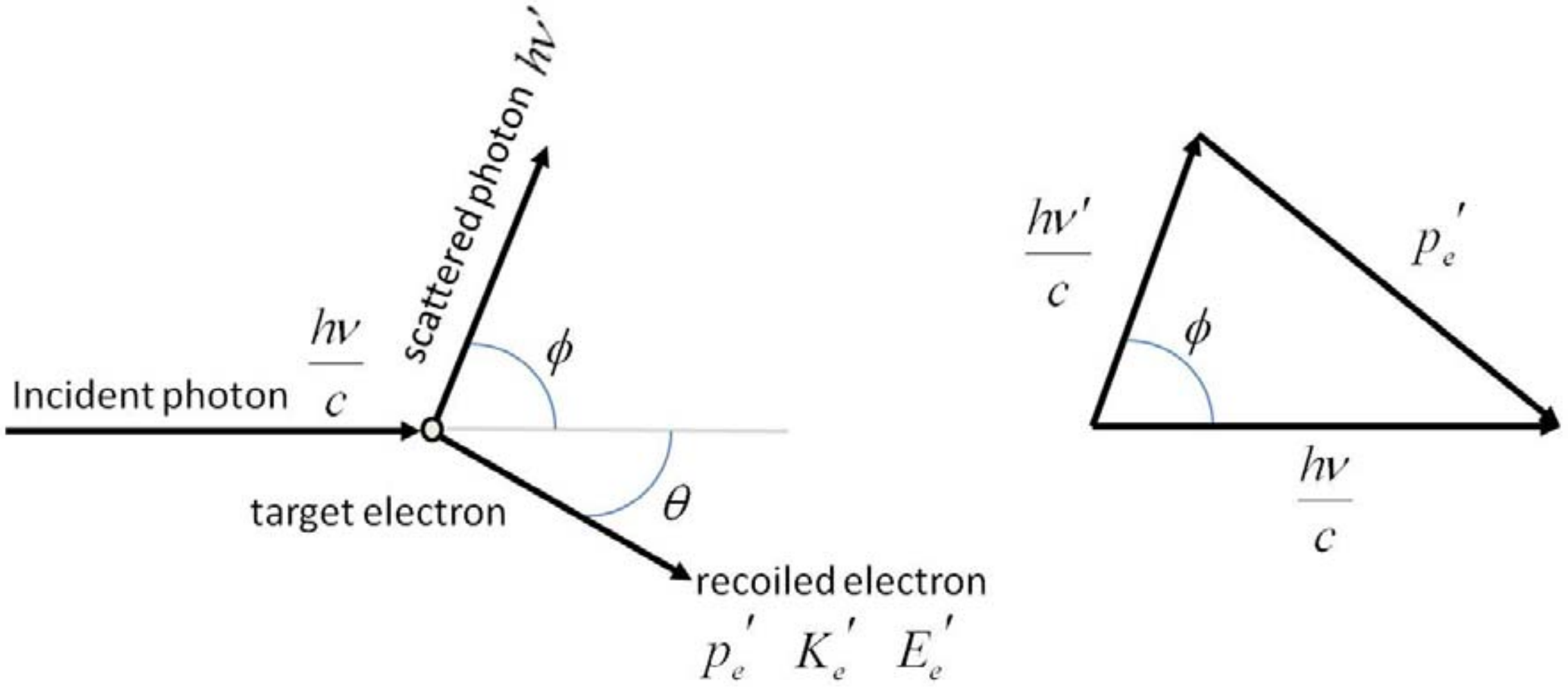

**Suppl. Fig. 1.** The geometry of Compton scattering showing the directions of scattered photon and recoiled electron with respect to the direction of the incident photon is shown in the left hand side, and conservation of momentum on the right hand side.

**The relation between momentum and kinetic energy of recoiled electrons.**

The momentum of the recoiled electron $p_e'(\phi)$ is related to its relativistic recoil kinetic energy $K_e'(\phi)$

$$p_e'(\phi) = c\left(\frac{K_e'(\phi)}{c^2} + m_0\right)\left(1 - \frac{1}{\left(K_e'(\phi)/m_0c^2\right)+1}\right)^{1/2} \tag{S1}$$

while assuming an initial stationary free electron.

**Suppl. Table 1**. The variation of the 'amplitude' of $KN(\phi)$ and $h\nu'(\phi)$ at $\phi = 0$, and of $p_e'(\phi)$ and $K_e'(\phi)$ at $\phi = \pi$ as a function of the energy of the incident photon. For incident photon energies less than 1 MeV, the $\pi$ in $h\nu'(\pi)$, $p_e'(\pi)$, and $K_e'(\pi)$ are replaced by $\phi$ at which the corresponding $KN(\phi)$ is minimum.

| $h\nu$ (MeV) | $\frac{(KN(0)-KN(\phi)_{\min})}{(KN(0)-KN(\phi)_{\min})_{1\text{GeV}}}$ | $\frac{(h\nu'(0)-h\nu'(\pi))}{h\nu}$ | $\left(\frac{p_{ph}}{p_e'(\pi)}\right)$ | $\left(\frac{K_e'(\pi)}{h\nu}\right)$ |
|---|---|---|---|---|
| 0.00001 | 0.5001 | $1.9495\times10^{-5}$ | 0.7084 | $1.9495\times10^{-5}$ |
| 0.0001 | 0.5003 | $1.9354\times10^{-4}$ | 0.7111 | $1.9354\times10^{-4}$ |
| 0.001 | 0.5020 | 0.0020 | 0.7040 | 0.0020 |
| 0.01 | 0.5190 | 0.0195 | 0.7077 | 0.0195 |
| 0.1 | 0.6457 | 0.1816 | 0.7276 | 0.1816 |
| 0.5 | 0.82626 | 0.5731 | 0.8167 | 0.5731 |
| 1 | 0.8941 | 0.7965 | 0.8309 | 0.7965 |
| 5 | 0.9758 | 0.9514 | 0.9536 | 0.9514 |
| 10 | 0.9877 | 0.9751 | 0.9757 | 0.9751 |
| 50 | 0.9976 | 0.9949 | 0.9949 | 0.9949 |
| 100 | 0.9988 | 0.9975 | 0.9975 | 0.9975 |
| 300 | 0.9997 | 0.9991 | 0.9991 | 0.9991 |
| 500 | 0.9999 | 0.9995 | 0.9995 | 0.9995 |
| 1000 | 1 | 0.9997 | 0.9997 | 0.9997 |

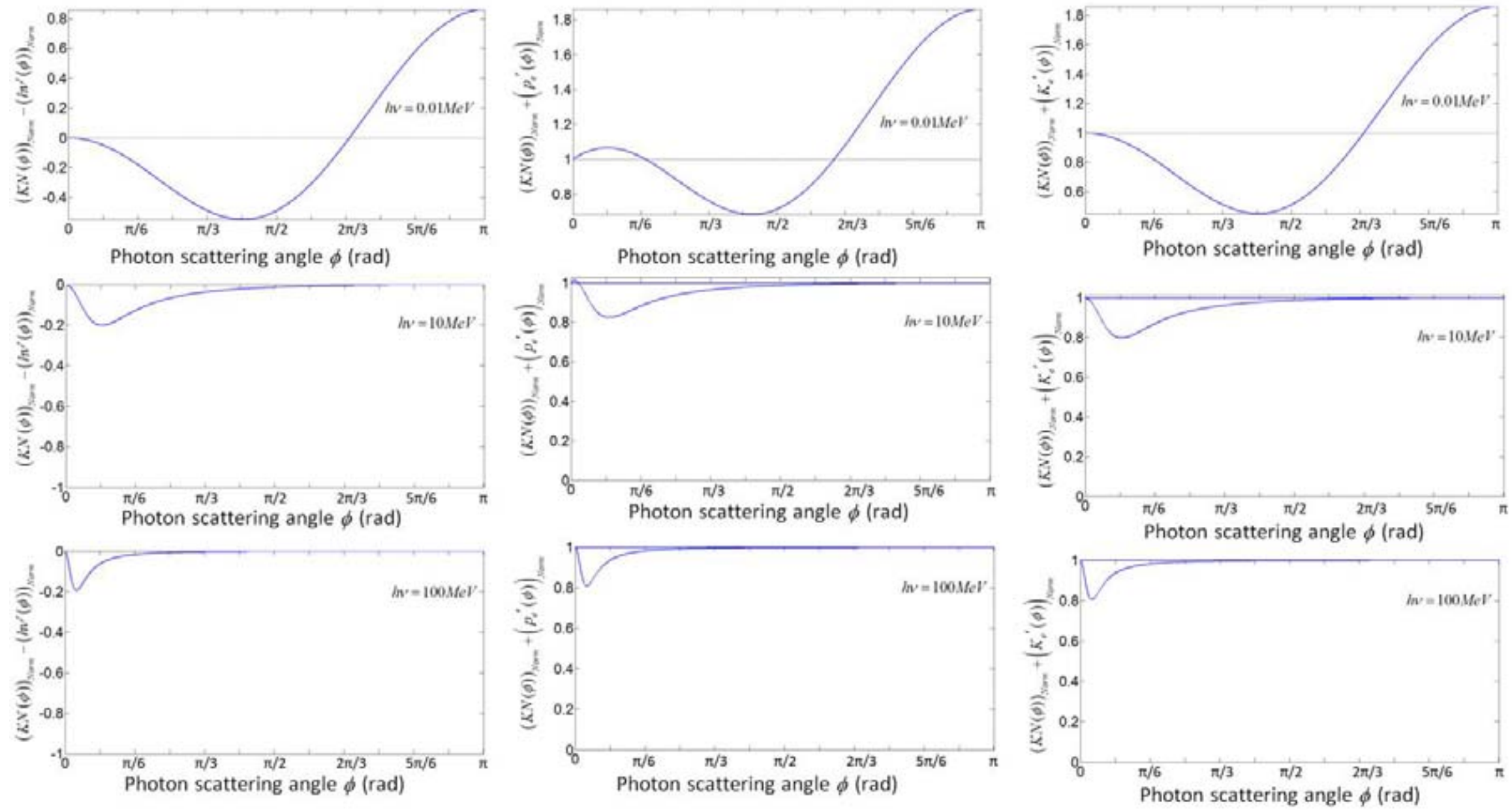

**Suppl. Fig. 2.** The variations of $\left[\left(KN(\phi)\right)_{Norm} - \left(h\nu'(\phi)\right)_{Norm}\right]$, $\left[\left(KN(\phi)\right)_{Norm} + \left(p_e{}'(\phi)\right)_{Norm}\right]$, and $\left[\left(KN(\phi)\right)_{Norm} + \left(K_e{}'(\phi)\right)_{Norm}\right]$ for incident photon energies are 0.01 MeV, 10 MeV, and 100 MeV, as a function of photon scattering angle $\phi$.

**Suppl. Table 2**. Pearson's chi-square test to compare the matching of $\left[\left(KN(\phi_i)\right)_{Norm}\right]$ with $\left[\left(h\nu'(\phi_i)\right)_{Norm}\right]$, and matching of $\left[\left(KN(\phi_i)\right)_{Norm}+\left(p_e{}'(\phi_i)\right)_{Norm}\right]$ and $\left[\left(KN(\phi_i)\right)_{Norm}+\left(K_e{}'(\phi_i)\right)_{Norm}\right]$ with 1 as a function of energy of incident photon $h\nu$ by considering 2000 data points (degree of freedom $=1999$), where a probability level of 0.9999 has a critical value of 1740.7049.

| $h\nu$ (MeV) | $\chi^2$ $O_i=\left[\left(KN(\phi_i)\right)_{Norm}\right]$ $E_i=\left[\left(h\nu'(\phi_i)\right)_{Norm}\right]$ | $\chi^2$ $O_i=\left[\left(KN(\phi_i)\right)_{Norm}+\left(p_e{}'(\phi_i)\right)_{Norm}\right]$ $E_i=1$ | $\chi^2$ $O_i=\left[\left(KN(\phi_i)\right)_{Norm}+\left(K_e{}'(\phi_i)\right)_{Norm}\right]$ $E_i=1$ |
|---|---|---|---|
| .00001 | 499.3880 | 391.2637 | 499.3880 |
| 0.0001 | 498.3787 | 390.2381 | 498.3787 |
| 0.001 | 488.4664 | 380.1696 | 488.4664 |
| 0.01 | 405.0711 | 295.7840 | 405.0711 |
| 0.1 | 154.8936 | 57.4931 | 154.8936 |
| 0.5 | 71.0068 | 18.5619 | 71.0068 |
| 1 | 44.6149 | 13.6493 | 44.6149 |
| 5 | 15.0225 | 9.1691 | 15.0225 |
| 10 | 10.1307 | 7.6830 | 10.1307 |
| 50 | 4.3626 | 4.1032 | 4.3626 |
| 100 | 3.0718 | 2.9777 | 3.0718 |
| 300 | 1.7690 | 1.7505 | 1.7690 |
| 500 | 1.3696 | 1.3610 | 1.3696 |
| 1000 | 0.9681 | 0.9651 | 0.9681 |

**Suppl. Table 3**. Pearson's chi-square test to compare the matching of $\left[\left(KN\left(\phi_i\right)\right)_{Norm}\right]$ with $\left[\left(h\nu'\left(\phi_i\right)\right)_{Norm}\right]$, and matching of $\left[\left(KN(\phi_i)\right)_{Norm}+\left(p_e'\left(\phi_i\right)\right)_{Norm}\right]$ and $\left[\left(KN(\phi_i\right)_{Norm}+\left(K_e'\left(\phi_i\right)\right)_{Norm}\right]$ with 1 as a function of energy of incident photon $h\nu$ by considering 2000 data points (degree of freedom $=1999$), where a probability level of 0.9999 has a critical value of 1740.7049, when $\sin^2\phi=0$ in $\left[\left(KN\left(\phi_i\right)\right)_{Norm}\right]$.

| $h\nu$ (MeV) | $\chi^2$ $O_i=\left[\left(KN\left(\phi_i\right)\right)_{Norm}\right]$ $E_i=\left[\left(h\nu'\left(\phi_i\right)\right)_{Norm}\right]$ | $\chi^2$ $O_i=\left[\left(KN(\phi_i)\right)_{Norm}+\left(p_e'\left(\phi_i\right)\right)_{Norm}\right]$ $E_i=1$ | $\chi^2$ $O_i=\left[\left(KN(\phi_i\right)_{Norm}+\left(K_e'\left(\phi_i\right)\right)_{Norm}\right]$ $E_i=1$ |
|---|---|---|---|
| .00001 | $4.0388\times10^{-8}$ | 52.3442 | $4.0388\times10^{-8}$ |
| 0.0001 | $4.0367\times10^{-6}$ | 52.3172 | $4.0367\times10^{-6}$ |
| 0.001 | $4.015\times10^{-4}$ | 52.0489 | $4.015\times10^{-4}$ |
| 0.01 | 0.0381 | 49.4488 | 0.0381 |
| 0.1 | 2.3281 | 30.2334 | 2.3281 |
| 0.5 | 12.1924 | 5.2747 | 12.1924 |
| 1 | 14.9837 | 2.3277 | 14.9837 |
| 5 | 11.5038 | 6.4720 | 11.5038 |
| 10 | 8.8197 | 6.5432 | 8.8197 |
| 50 | 4.2392 | 3.9832 | 4.2392 |
| 100 | 3.0278 | 2.9343 | 3.0278 |
| 300 | 1.7604 | 1.7420 | 1.7604 |
| 500 | 1.3656 | 1.3570 | 1.3656 |
| 1000 | 0.9667 | 0.9637 | 0.9667 |